# How to test the "quantumness" of a quantum computer?

*Alexandre M. Zagoskin[1,2], Evgeni Il'ichev[3], Miroslav Grajcar[4], Joseph J. Betouras[1], Franco Nori[2,5]*

[1]Physics Department, Loughborough University, Loughborough LE11 3TU, UK
[2]Center for Emergent Matter Science, RIKEN, Saitama 351-0198, Japan
[3]Quantum Detection, Institute of Photonic Technology, 07702 Jena, Germany
[4]Department of Experimental Physics, Comenius University, 811 02 Bratislava, Slovakia
[5]Department of Physics, University of Michigan, Ann Arbor, 48109, USA

Suppose we are given a black box and told that this is a quantum computer. How do we test if it is capable of performing quantum computing? The standard approach would be to run a series of tests of few-qubit operations, and compare the outputs with results obtained using a standard (classical) computer simulating the quantum evolution. This is possible for small-size systems, likely comprising less than 40-50 qubits, given the current capabilities and fundamental limitations of classical computers (due to the exponential growth of computing resources needed to model large quantum systems [1]). But what if the system size is larger? Another challenging task would be to find out how to change the design of the "black-box computer" to improve its performance. Namely, if the black box computer is *not* working quantum mechanically, how to find this out and how to fix the problem?

Now we are facing a similar situation. The development of an "adiabatic quantum annealer" by D-Wave Systems Inc. over the last few years did produce reactions ranging from excitement to skepticism [2-7]. Considerable interest was generated by:

(1) the demonstration of quantum annealing in an 8-qubit register of the prototype processor [8];
(2) the sale of a 128-qubit quantum annealer "D-Wave One" to Lockheed Martin and its installation at the University of Southern California (2011), and its upgrade to a 512-qubit "D-Wave Two" (2013);
(3) the subsequent evidence of quantum annealing in the working 108 qubits of this device [9];
(4) the realization of a quantum adiabatic algorithm on a nominally 512-qubit (439 qubits operational) device "D-Wave Two" outperforming at least some classical algorithms [10];
(5) the May 2013 decision by NASA, Google and the Universities Space Research Association to purchase a "D-Wave Two" to be installed at their common Quantum Artificial Intelligence Lab at NASA's Ames Research Centre [11];
(6) the evaluation of small Ramsey numbers using the "D-Wave One" device [12].

Questions were raised concerning the relative speed of the D-Wave processors compared to classical optimization algorithms [7] and the quantum character of their evolution (see [13] and the response [14]). While the former question is crucial for information science, the latter one is more important from the point of view of physics. We believe that resolving it would drastically improve the current status of the field, which we find unsatisfactory in several important respects. The very fundamental impossibility of an efficient simulation, with classical means, of large enough quantum systems [1], which provides the justification for the quest for quantum computing, may prevent us from developing quantum computers, unless better classical approaches towards their design and evaluation are found.

## Current situation

The device ("D-Wave One"), tested in [9], consisted of 128 (108 of which were operational) Nb-based superconducting flux qubits (see, e.g., [15], Ch.2), arranged in blocks of 8, with selective tunable couplings. This particular design used dc SQUIDs in place of Josephson junctions, which allowed to fine-tune the qubit parameters by changing the magnetic fluxes through the control loops. The amplitude and sign of the couplings between the qubits could also be tuned in a similar way [16]. The qubits formed a lattice, which can be modeled by a network of Ising spins with randomly chosen interactions.

The goal of the experiment [9] was to find the statistics of the device by determining the random spin glass ground state and comparing the results with the algorithms based on simulated classical and quantum annealing (SCA and SQA, respectively). The surprising result was that the operation of D-Wave One produced a bimodal statistical distribution of success probabilities, corresponding to clearly distinct groups of "easy" and "hard" problems, similar to the one produced by SQA, but drastically different from SCA. In addition, they detected strong positive correlations of the success probabilities at different instances between the D-Wave One processor and SQA. As pointed out in [13], the bimodality itself was not sufficient to claim the evidence for quantum annealing, as it could be reproduced by semi-classical spin models. Nevertheless, the absence of correlations between semi-classical models and D-Wave One made a strong case in favor of quantum behavior, demonstrating some essential features of quantum annealing [9,14]. The above results are surprising, since the adiabatic evolution time of the processor (5-15 μs) greatly exceeded the decoherence time of each separate qubit (~100 ns) and therefore of the processor as a whole.

## Quantum coherence and entanglement

While the critical importance of maintaining quantum coherence for gate-based quantum computing is firmly established, the question of its role for universal adiabatic quantum computing, and its more limited versions (such as quantum optimization or approximate adiabatic quantum computing) is being debated (see, e.g., [17,18]). Quantum coherence is certainly necessary, but on what scale, and for how long?

There is a point of view, according to which the existence of entangled energy eigenstates is not only necessary, but it may even be sufficient, for at least a limited operation of an AQC. This was guiding some early efforts in this field [19]. Due to the fact that the tested structures consisted of a *small number* of superconducting flux qubits, it was possible to establish the existence of such eigenstates by the direct modeling of the quantum evolution of the system, and its comparison with the experimental data was possible. But for a general system the question remains: multiqubit entanglement is certainly necessary, but on what scale, and for how long?

## Scaling with the number of qubits

The situation becomes qualitatively different with the current and future generations of multiqubit processors. The simulation of the full 108-qubit system in the quantum limit was not attempted, and therefore a direct investigation of the role of entangled energy eigenstates could not be undertaken. (Difficulties of such simulation were already explicit in [8], where only a single 8-qubit register was investigated.) It was only possible to make a conjecture [14] that the better correlations between the behavior of the D-Wave processor and the SQA (compared to semi-classical spin models) was due to the fact that entangled energy eigenstates were used in the SQA calculations, but not in the semiclassical models. The exact classical optimization algorithms used in [9], as well as generic approximate algorithms, will take impractically long time to run for 512 qubits. For the specific structure of the D-Wave One processor, it is expected that the simulated annealer algorithms (both classical and quantum) will scale exponentially (as $\exp[a\sqrt{N}]$) with the number *N* of qubits, the same as exact solvers [9]. Therefore, it is likely that their usefulness for establishing the quantum character of their evolution in the planned devices, with several thousand qubits, is questionable. However, since SQA does not simulate the actual quantum behaviour of the system, its results are of questionable relevance for the prediction of the ability of a given large-scale quantum qubit array to demonstrate quantum behaviour at a given level of environmental and intrinsic decoherence, dispersion of parameters, etc.

## The real problem and possible solutions

The looming impossibility to predict the behaviour of *any* big enough quantum processor (adiabatic, gate-based, etc) and even to test it for "quantumness" using classical tools, is the elephant in the room, and it may effectively restrict any further progress. Even taking the optimistic view, that quantum computing is not fundamentally restricted (by, e.g., limits on the size of systems capable of demonstrating quantum behaviour [20]), it is realistic to expect, based on the current state of art, that a quantum processor capable of simulating itself accurately and quickly enough to be useful, will contain significantly more qubits than the current or prospective D-Wave machines.

A recent analysis of the perspectives of superconducting circuits as a platform for universal quantum computation [21] stressed the very high price of implementing quantum error correction (necessary for the gate-based quantum computing) or "surface code" quantum computing [22], which runs to hundreds or thousands of physical qubits per a logical qubit. Reference [21] speculates that large quantum processors should perhaps rely on a modular approach (when the operation and functionality of unit modules can be separately tested and characterized), or on some "hardware – specific shortcuts" (like using nonlinear oscillators instead of qubits as a basis for superconducting quantum computing). However, these speculations may be overly optimistic. In order to use quantum parallelism one should entangle a few dozen logical qubits, which is only possible if all, or a significant fraction, of the unit elements are in a quantum coherent state for some minimum period of time. It was recently demonstrated experimentally [23], that the quantumness of a gate-based quantum computer can be verified using a smaller quantum device. It would be very interesting to know whether this approach can be extended to quantum annealers or used to estimate the performance of such devices. But so far, we cannot avoid the need to estimate and evaluate the behaviour of large, essentially quantum systems with *classical* means, in order to develop a useful quantum computer.

The problem pertains for quantum annealers, universal quantum computers (gate-based or otherwise), adiabatic quantum computers, quantum simulators, and even for simpler artificial quantum structures, such as quantum metamaterials, whose properties are essentially determined by quantum correlations and entanglement within large collections of artificial atoms interacting with the electromagnetic field.

Nevertheless, we believe that this problem can be solved. This will first require developing a better set of theoretical tools. A system of qubits is, after all, a quantum many-body system, which may be amenable to the approaches which worked so well in many applications to condensed matter physics and statistical mechanics. If, for the moment, we restrict the field to quantum annealers, then the existing theoretical formalism must be extended to efficiently include the two essential features of this problem: its essentially nonequilibrium, transitional character, and the importance of quantum coherence (e.g., following [24]). It would be desirable to have an efficient method of establishing the probability, that the observed set of runs of a large-enough quantum annealer cannot be reduced to classical physics, and of estimating the performance of such a device based solely on the device parameters. This would not require to classically simulate a particular run of the device.

Such set of theoretical tools will be useful not just for applications to quantum computing, but in a wider field of quantum engineering and second-generation quantum technologies [25]. But the whole area cannot flourish until the proper guidelines are obtained on what kind of systems and under what conditions macroscopic quantum behavior is likely to be realized, and what should be the signs of this realization. Obviously, how to evaluate the "quantumness" of a black box is a challenging problem, which requires a concerted approach.